\documentclass[pre,twocolumn]{revtex4-1}
\usepackage{amsmath}
\usepackage{latexsym}
\usepackage{amssymb}
\usepackage{mathrsfs}
\usepackage{bm}
\usepackage{indentfirst}
\usepackage{natbib}
\usepackage{graphicx}
\usepackage{relsize}
\usepackage{epstopdf}
\usepackage{comment}
\usepackage{lmodern}
\usepackage[T1]{fontenc}
\usepackage{verbatim}
\usepackage{enumerate}
\usepackage{setspace}
\usepackage{color}
\usepackage{xcolor}

\begin{document}

 \title{Regularization and decimation pseudolikelihood approaches to statistical inference in $XY$-spin models}
 
\author{Payal Tyagi$^{1,2}$, Alessia Marruzzo$^{2}$, Andrea Pagnani$^{3,4}$, Fabrizio Antenucci$^{2}$, Luca Leuzzi$^{2,1}$}

\address{
$^1$ Dipartimento di Fisica, Universit\`a {\em Sapienza},
 Piazzale Aldo Moro 5, I-00185, Rome, Italy }
 \address{
 $^2$  CNR-NANOTEC, Institute of Nanotechnology, Rome - {\em Soft and Living
  Matter Lab.},
  Piazzale Aldo Moro 5, I-00185, Rome, Italy }
  \address{
   $^3$ Department of Applied Science and Technology and Center for Computational Sciences, Politecnico di Torino, Corso Duca degli Abruzzi 24, Torino, Italy}
\address {$^4$ Human Genetics Foundation-Torino, Via Nizza 52, Torino, Italy}

\begin{abstract}
We implement  a pseudolikelyhood approach with $l_2$-regularization as well as the recently introduced pseudolikelihood with decimation procedure to the inverse problem in continuous spin models on arbitrary networks, with arbitrarily disordered couplings. Performances of the approaches are tested against data produced by Monte Carlo numerical simulations and compared also from previously studied fully-connected mean-field-based inference techniques. The results clearly show that the best network reconstruction is obtained through the decimation scheme, that also allows to dwell the inference down to lower temperature regimes. Possible applications to phasor models for light propagation in random media are proposed and discussed. 
\end{abstract} 

\maketitle
\section{Introduction}
Given a data set and a model with some unknown parameters, the inverse problem aims to find the values of the model parameters that best fit the data.  
In this work, in which we focus on systems of interacting elements,
 the inverse problem concerns  the statistical inference
 of the underling interaction network and of its coupling coefficients from observed data on the dynamics  of the system. 
 Versions of this problem are encountered in physics, biology (e.g., \cite{Balakrishnan11,Ekeberg13,Christoph14}), social sciences and finance (e.g.,\cite{Mastromatteo12,yamanaka_15}), neuroscience (e.g., \cite{Schneidman06,Roudi09a,tyrcha_13}), just to cite a few, and are becoming more and more important due to the increase in the amount of data available from these fields.\\
 \indent 
 A standard approach used in statistical inference is to predict the interaction couplings by maximizing the likelihood function.
 This technique, however, requires the evaluation of the  
   partition function that, in the most general case, concerns a number of computations scaling exponentially with the system size.
    Boltzmann machine learning uses Monte Carlo sampling to compute the gradients of the Log-likelihood looking for stationary points \cite{Murphy12} but this method is computationally manageable only for small systems. A series of faster approximations, such as naive mean-field, independent-pair approximation \cite{Roudi09a, Roudi09b}, inversion of TAP equations \cite{Kappen98,Tanaka98}, small correlations expansion \cite{Sessak09}, adaptive TAP \cite{Opper01}, adaptive  cluster expansion \cite{Cocco12} or Bethe approximations \cite{Ricci-Tersenghi12, Nguyen12}  have, then, been developed. These techniques take as input means and correlations of observed variables and most of them assume a fully connected graph as underlying connectivity network, or expand  around it by perturbative dilution.  In most cases, network reconstruction turns out to  be not accurate for small data sizes and/or when couplings are  strong or, else, if the original interaction network is sparse.\\
\indent
 A further method, substantially improving performances for small data, is the so-called Pseudo-Likelyhood Method (PLM) \cite{Ravikumar10}. In Ref. \cite{Aurell12} Aurell and Ekeberg performed a comparison between PLM and some of the just mentioned mean-field-based algorithms on the pairwise interacting Ising-spin  ($\sigma = \pm 1$) model, showing how PLM performs sensitively better, especially on sparse graphs and in the high-coupling limit, i.e., for low temperature.
 
 In this work, we aim at performing statistical inference  on a model whose interacting variables are continuous $XY$ spins, i.e., $\sigma \equiv \left(\cos \phi,\sin \phi\right)$ with $\phi \in [0, 2\pi )$. The developed tools can, actually, be also straightforward applied  to the $p$-clock model  \cite{Potts52} where the phase $\phi$ takes discretely equispaced $p$ values  in the $2 \pi$ interval, $\phi_a =  a 2 \pi/p$, with $a= 0,1,\dots,p-1$. The $p$-clock model, else called vector Potts model, gives a hierarchy of discretization of the $XY$ model as $p$ increases. For $p=2$, one recovers the Ising model, for $p=4$ the Ashkin-Teller model \cite{Ashkin43}, for $p=6$ the ice-type model \cite{Pauling35,Baxter82} and the eight-vertex model \cite{Sutherland70,Fan70,Baxter71} for $p=8$.  
It turns out to be very useful also for numerical implementations of the continuous $XY$ model. 
Recent analysis on the multi-body $XY$ model has shown that for a limited number of discrete phase values ($p\sim 16, 32$) the thermodynamic critical properties of the $p\to\infty$ $XY$ limit are promptly recovered \cite{Marruzzo15, Marruzzo16}.  
Our main motivation to study statistical inference is that these kind of models have recently turned out to be rather useful in describing the behavior of optical systems, 
including standard mode-locking lasers \cite{Gordon02,Gat04,Angelani07,Marruzzo15} and random lasers \cite{Angelani06a,Leuzzi09a,Antenucci15a,Antenucci15b,Marruzzo16}. 
In particular, the inverse problem on the pairwise XY model analyzed here might be of help in recovering images from light propagated through random media.

      This paper is organized as follows: in Sec. \ref{sec:model} we introduce the general model  and we discuss its  derivation also  as a model for light transmission through random scattering media. 
    In Sec. \ref{sec:plm} we introduce the PLM with $l_2$ regularization  and with decimation, two variants of the PLM respectively introduced in Ref. \cite{Wainwright06} and \cite{Aurell12} for the inverse Ising problem. 
    Here, we analyze these techniques for continuous $XY$ spins and we test them on thermalized data generated by Exchange Monte Carlo numerical simulations of the original model dynamics. In Sec. \ref{sec:res_reg} we  present the results related to the PLM-$l_2$. In Sec. \ref{sec:res_dec} the results related to the PLM with decimation are reported and its performances are compared to the PLM-$l_2$ and to a variational mean-field method analyzed in Ref. \cite{Tyagi15}. In Sec. \ref{sec:conc}, we outline conclusive remarks and perspectives.

        \section{The leading $XY$ model}
      \label{sec:model}
 The leading model we are considering is defined, for a system of $N$ angular $XY$ variables, by the Hamiltonian 
 \begin{equation}
  \mathcal{H} = - \sum_{ik}^{1,N} J_{ik} \cos{\left(\phi_i-\phi_k\right)} 
  \label{eq:HXY}
  \end{equation} 
  
  The $XY$ model is well known in statistical mechanics, displaying important physical
  insights, starting from the Berezinskii-Kosterlitz-Thouless
  transition in two dimensions\cite{Berezinskii70,Berezinskii71,Kosterlitz72} and moving to, e.g., the
  transition of liquid helium to its superfluid state \cite{Brezin82}, the roughening transition of the interface of a crystal in equilibrium with its vapor \cite{Cardy96}. In presence of disorder and frustration \cite{Villain77,Fradkin78} the model has been adopted to describe synchronization problems as the Kuramoto model \cite{Kuramoto75} and in the theoretical modeling of Josephson junction arrays \cite{Teitel83a,Teitel83b} and arrays of coupled lasers \cite{Nixon13}.
  Besides several derivations and implementations of the model in quantum and classical physics, equilibrium or out of equilibrium, ordered or fully frustrated systems, Eq. (\ref{eq:HXY}), in its generic form,
  has found applications also in other fields. A rather fascinating example being the behavior of starlings flocks  \cite{Reynolds87,Deneubourg89,Huth90,Vicsek95, Cavagna13}.
  Our interest on the $XY$ model resides, though, in optics. Phasor and phase models with pairwise and  multi-body interaction terms can, indeed, describe the behavior of electromagnetic modes in both linear and nonlinear optical systems in the analysis of problems such as light propagation and lasing \cite{Gordon02, Antenucci15c, Antenucci15d}. As couplings are strongly frustrated, these models turn out to be especially useful to the study of optical properties in random media \cite{Antenucci15a,Antenucci15b}, as in the noticeable case of random lasers \cite{Wiersma08,Andreasen11,Antenucci15e}  and they might as well be applied to linear scattering problems, e.g., propagation of waves in opaque systems or disordered fibers.

  \subsection{A propagating wave model}
  We briefly mention a derivation of the model as a proxy for the propagation of light through random linear media. 
 Scattering of light is held responsible to obstruct our view and make objects opaque. Light rays, once that they enter the material, only exit  after getting scattered multiple times within the material. In such a disordered medium, both the direction and the phase of the propagating waves are random. Transmitted light 
 yields a disordered interference pattern typically  having low intensity, random phase and almost no resolution, called a speckle. Nevertheless, in recent years it has been realized that disorder is rather a blessing in disguise \cite{Vellekoop07,Vellekoop08a,Vellekoop08b}. Several experiments have made it possible to control the behavior of light and other optical processes in a given random disordered medium,  
 by exploiting, e.g.,  the tools developed for wavefront shaping to control the propagation of light and to engineer the confinement of light \cite{Yilmaz13,Riboli14}.
 \\
 \indent
 In a linear dielectric medium, light propagation can be described through a part of the scattering matrix, the transmission matrix $\mathbb{T}$, linking the outgoing to the incoming fields. 
 Consider the case in which there are $N_I$ incoming channels and $N_O$ outgoing ones; we can indicate with $E^{\rm in,out}_k$ the input/output electromagnetic field phasors of channel $k$. In the most general case, i.e., without making any particular assumptions on the field polarizations, each light mode and its polarization polarization state can be represented by means of the $4$-dimensional Stokes vector. Each $ t_{ki}$ element of $\mathbb{T}$, thus, is a $4 \times 4$ M{\"u}ller matrix. If, on the other hand, we know that the source is polarized and the observation is made on the same polarization, one can use a scalar model and adopt Jones calculus \cite{Goodman85,Popoff10a,Akbulut11}:
   \begin{eqnarray}
 E^{\rm out}_k = \sum_{i=1}^{N_I}  t_{ki} E^{\rm in}_i \qquad \forall~ k=1,\ldots,N_O
 \label{eq:transm}
 \end{eqnarray}
  We recall that the elements of the transmission matrix are random complex coefficients\cite{Popoff10a}. For the case of completely unpolarized modes, we can also use a scalar model similar to Eq. \eqref{eq:transm}, but whose variables are  the intensities of the outgoing/incoming fields, rather than the fields themselves.\\ 
In the following, for simplicity, we will consider Eq. (\ref{eq:transm}) as our starting point,
where $E^{\rm out}_k$, $E^{\rm in}_i$ and $t_{ki}$ are all complex scalars. 
If Eq. \eqref{eq:transm} holds for any $k$, we can write:
  \begin{eqnarray}
  \int \prod_{k=1}^{N_O} dE^{\rm out}_k \prod_{k=1}^{N_O}\delta\left(E^{\rm out}_k - \sum_{j=1}^{N_I}  t_{kj} E^{\rm in}_j \right) = 1
  \nonumber
  \\
  \label{eq:deltas}
  \end{eqnarray}

 Observed data are a noisy representation of the true values of the fields. Therefore, in inference problems it is statistically more meaningful to take that noise into account in a probabilistic way, 
 rather than looking  at the precise solutions of the exact equations (whose parameters are unknown). 
 To this aim we can introduce Gaussian distributions whose limit for zero variance are the Dirac deltas in Eq. (\ref{eq:deltas}).
 Moreover, we move to consider the ensemble of all possible solutions of Eq. (\ref{eq:transm}) at given $\mathbb{T}$, looking at  all configurations of input fields. We, thus, define the function:
 
   \begin{eqnarray}
  Z &\equiv &\int_{{\cal S}_{\rm in}} \prod_{j=1}^{N_I}  dE^{\rm in}_j \int_{{\cal S}_{\rm out}}\prod_{k=1}^{N_O} dE^{\rm out}_k 
  \label{def:Z}
\\
    \times
  &&\prod_{k=1}^{N_O}
   \frac{1}{\sqrt{2\pi \Delta^2}}  \exp\left\{-\frac{1}{2 \Delta^2}\left|
  E^{\rm out}_k -\sum_{j=1}^{N_I}  t_{kj} E^{\rm in}_j\right|^2
\right\} 
\nonumber
 \end{eqnarray}
   We stress that the integral of Eq. \eqref{def:Z} is not exactly a Gaussian integral. Indeed, starting from Eq. \eqref{eq:deltas}, two constraints on the electromagnetic field intensities must be taken into account. 
  The space of solutions is  delimited by the total power ${\cal P}$ received by system, i.e., 
  ${\cal S}_{\rm in}: \{E^{\rm in} |\sum_k I^{\rm in}_k = \mathcal{P}\}$, also implying  a constraint on the total amount of energy that is transmitted through the medium, i. e., 
  ${\cal S}_{\rm out}:\{E^{\rm out} |\sum_k I^{\rm out}_k=c\mathcal{P}\}$, where the attenuation factor  $c<1$ accounts for total losses.
  As we will see more in details in the following, being interested in inferring the transmission matrix through the PLM, we can omit to explicitly include these terms in Eq. \eqref{eq:H_J} since they do not depend on $\mathbb{T}$ not adding any information on the gradients with respect to the elements of $\mathbb{T}$.
  
 Taking the same number of incoming and outcoming channels, $N_I=N_O=N/2$, and  ordering the input fields in the first $N/2$ mode indices and the output fields in the last $N/2$ indices, we can drop the ``in'' and ``out'' superscripts and formally write $Z$  as a partition function
    \begin{eqnarray}
        \label{eq:z}
 && Z =\int_{\mathcal S} \prod_{j=1}^{N} dE_j \left(   \frac{1}{\sqrt{2\pi \Delta^2}} \right)^{N/2} 
 \hspace*{-.4cm} \exp\left\{
  -\frac{ {\cal H} [\{E\};\mathbb{T}] }{2\Delta^2}
  \right\}
  \\
&&{\cal H} [\{E\};\mathbb{T}] =
-  \sum_{k=1}^{N/2}\sum_{j=N/2+1}^{N} \left[E^*_j t_{jk} E_k + E_j t^*_{kj} E_k^* 
\right]
 \nonumber
\\
&&\qquad\qquad \qquad + \sum_{j=N/2+1}^{N} |E_j|^2+ \sum_{k,l}^{1,N/2}E_k
U_{kl} E_l^*
 \nonumber
 \\
 \label{eq:H_J}
 &&\hspace*{1.88cm } = - \sum_{nm}^{1,N} E_n J_{nm} E_m^*
 \end{eqnarray}
 where ${\cal H}$ is a real-valued function by construction, we have introduced the effective input-input coupling matrix
\begin{equation}
U_{kl} \equiv \sum_{j=N/2+1}^{N}t^*_{lj} t_{jk} 
 \label{def:U}
 \end{equation}
 and the whole interaction matrix reads (here $\mathbb{T} \equiv \{ t_{jk} \}$)
 \begin{equation}
 \label{def:J}
 \mathbb J\equiv \left(\begin{array}{ccc|ccc}
 \phantom{()}&\phantom{()}&\phantom{()}&\phantom{()}&\phantom{()}&\phantom{()}\\
 \phantom{()}&-\mathbb{U} \phantom{()}&\phantom{()}&\phantom{()}&{\mathbb{T}}&\phantom{()}\\
\phantom{()}&\phantom{()}&\phantom{()}&\phantom{()}&\phantom{()}&\phantom{()}\\
 \hline
\phantom{()}&\phantom{()}&\phantom{()}&\phantom{()}&\phantom{()}&\phantom{()}\\
 \phantom{()}& \mathbb   T^\dagger&\phantom{()}&\phantom{()}& - \mathbb{I} &\phantom{()}\\
\phantom{a}&\phantom{a}&\phantom{a}&\phantom{a}&\phantom{a}&\phantom{a}\\
 \end{array}\right)
 \end{equation}
 
 Determining the electromagnetic complex amplitude configurations that minimize the {\em cost function} ${\cal H}$, Eq.  (\ref{eq:H_J}),  means to maximize the overall distribution peaked around the solutions of the transmission Eqs. (\ref{eq:transm}). As the variance $\Delta^2\to 0$, eventually, the initial set of Eqs. (\ref{eq:transm}) are recovered. The ${\cal H}$ function, thus, plays the role of an Hamiltonian and  $\Delta^2$ the role of a noise-inducing temperature. The exact numerical problem corresponds to the zero temperature limit of the statistical mechanical problem. Working with real data, though, which are noisy, a finite ``temperature''
  allows for a better representation of the ensemble of solutions to the sets of equations of continuous variables.

  Now, we can express every phasor in Eq. \eqref{eq:z}  as $E_k = A_k e^{\imath \phi_k}$. As a working hypothesis we will consider the intensities $A_k^2$ as either homogeneous or as \textit{quenched} with respect to phases.
The first condition occurs, for instance, to the input intensities $|E^{\rm in}_k|$ produced by a phase-only spatial light modulator (SLM) with homogeneous illumination \cite{Popoff11}.
With \textit{quenched} here we mean, instead, that the intensity of each mode is the same for every solution of Eq. \eqref{eq:transm} at fixed $\mathbb T$.
We stress that, including intensities in the model does not preclude the inference analysis but it is out of the focus of the present work and will be considered elsewhere. 

If all intensities are uniform in input and in output, this amount to a constant rescaling for each one of the four sectors of matrix $\mathbb J$ in Eq. (\ref{def:J}) that will not change the properties of the matrices.
For instance, if the original transmission matrix is unitary, so it will be the rescaled one and the matrix $\mathbb U$ will be  diagonal.
Otherwise, if intensities are \textit{quenched}, i.e., they can be considered as constants in Eq. (\ref{eq:transm}),
they are inhomogeneous with respect to phases. The generic Hamiltonian element will, therefore, rescale as 
  \begin{eqnarray}
  E^*_n J_{nm} E_m = J_{nm} A_n A_m e^{\imath (\phi_n-\phi_m)} \to J_{nm} e^{\imath (\phi_n-\phi_m)}
  \nonumber
  \end{eqnarray}
  and the properties of the original  $J_{nm}$ components are not conserved  in the rescaled one. In particular, we have no argument, anymore, to possibly set the rescaled $U_{nm}\propto \delta_{nm}$.
  Eventually, we end up with the complex couplings $XY$ model, whose real-valued Hamiltonian is written as
 \begin{eqnarray}
  \mathcal{H}& = &  - \frac{1}{2} \sum_{nm} J_{nm} e^{-\imath (\phi_n - \phi_m)}  + \mbox{c.c.} 
    \label{eq:h_im}
\\    &=&  - \frac{1}{2} \sum_{nm} \left[J^R_{nm} \cos(\phi_n - \phi_m)+
  J^I_{nm}\sin (\phi_n - \phi_m)\right] 
  \nonumber
 \end{eqnarray}
where $J_{nm}^R$ and $J_{nm}^I$ are the real and imaginary parts of $J_{nm}$. Being $\mathbb J$  Hermitian, $J^R_{nm}=J^R_{mn}$ is symmetric and $J_{nm}^I=-J_{mn}^I$ is skew-symmetric.

  \section{Pseudolikelihood Maximization}
  \label{sec:plm}
The inverse problem consists in the reconstruction of the parameters $J_{nm}$ of the Hamiltonian, Eq. (\ref{eq:h_im}).  
Given a set of $M$ data configurations of $N$ spins
 $\bm\sigma = \{ \cos \phi_i^{(\mu)},\sin \phi_i^{(\mu)} \}$, $i = 1,\dots,N$ and $\mu=1,\dots,M$, we want to \emph{infer} the couplings:
 \begin{eqnarray}
\bm \sigma  \rightarrow  \mathbb{J} 
\nonumber
 \end{eqnarray}
 With this purpose in mind,
 in the rest of this section we implement the working equations for the techniques used. 
 In order to test our methods, we generate the input data, i.e., the configurations, by Monte-Carlo simulations of the model.
 The joint probability distribution of the $N$ variables $\bm{\phi}\equiv\{\phi_1,\dots,\phi_N\}$, follows the Gibbs-Boltzmann distribution:
 \begin{equation}\label{eq:p_xy}
 P(\bm{\phi}) = \frac{1}{Z} e^{-\beta \mathcal{H\left(\bm{\phi}\right)}} \quad \mbox{ where } \quad Z = \int \prod_{k=1}^N d\phi_k  e^{-\beta \mathcal{H\left(\bm{\phi}\right)}}  
 \end{equation}
 and where we denote $\beta=\left( 2\Delta^2 \right)^{-1}$ with respect to Eq. (\ref{def:Z}) formalism.
 In order to stick to usual statistical inference notation, in the following we will rescale the couplings by a factor $\beta / 2$: $\beta J_{ij}/2 \rightarrow J_{ij}$. 
 The main idea of the PLM is to work with the conditional probability distribution of one variable $\phi_i$ given all other variables, 
 $\bm{\phi}_{\backslash i}$:
 
  \begin{eqnarray}
	\nonumber
   P(\phi_i | \bm{\phi}_{\backslash i}) &=& \frac{1}{Z_i} \exp \left \{ {H_i^x (\bm{\phi}_{\backslash i})
  	\cos \phi_i + H_i^y (\bm{\phi}_{\backslash i}) \sin \phi_i } \right \}
	\\
 \label{eq:marginal_xy}
	&=&\frac{e^{H_i(\bm{\phi}_{\backslash i}) \cos{\left(\phi_i-\alpha_i(\bm{\phi}_{\backslash i})\right)}}}{2 \pi I_0(H_i)}
  \end{eqnarray}
  where $H_i^x$ and $H_i^y$ are defined as
   \begin{eqnarray}
   H_i^x (\bm{\phi}_{\backslash i})  &=& \sum_{j (\neq i)} J^R_{ij} \cos \phi_j  - \sum_{j  (\neq i) } J_{ij}^{I} \sin \phi_j \phantom{+  h^R_i} \label{eq:26} \\
   H_i^y  (\bm{\phi}_{\backslash i})  &=&  \sum_{j  (\neq i)} J^R_{ij}  \sin \phi_j  + \sum_{j  (\neq i) } J_{ij}^{I}   \cos \phi_j \phantom{ + h_i^{I} }\label{eq:27}
   \end{eqnarray}
and $H_i= \sqrt{(H_i^x)^2 + (H_i^y)^2}$, $\alpha_i = \arctan H_i^y/H_i^x$   and we introduced the modified Bessel function of the first kind:
  \begin{equation}
  \nonumber
   I_k(x) = \frac{1}{2 \pi}\int_{0}^{2 \pi} d \phi e^{x \cos{ \phi}}\cos{k \phi}
  \end{equation}
  
   Given $M$ observation samples $\bm{\phi}^{(\mu)}=\{\phi^\mu_1,\ldots,\phi^\mu_N\}$, $\mu = 1,\dots, M$, the
   pseudo-loglikelihood for the variable $i$ is given by the logarithm of Eq. (\ref{eq:marginal_xy}),
   \begin{eqnarray}
   \label{eq:L_i}
   L_i &=& \frac{1}{M}  \sum_{\mu = 1}^M  \ln P(\phi_i^{(\mu)}|\bm{\phi}^{(\mu)}_{\backslash i})
   \\
   \nonumber
   & =&  \frac{1}{M}  \sum_{\mu = 1}^M  \left[ H_i^{(\mu)} \cos( \phi_i^{(\mu)} - \alpha_i^{(\mu)}) - \ln  2 \pi I_0\left(H_i^{(\mu)}\right)\right] \, .
    \end{eqnarray}
The underlying idea of PLM is that an approximation of the true parameters of the model is obtained for values that maximize the functions $L_i$.
The specific maximization scheme differentiates the different techniques.

   \subsection{PLM with $l_2$ regularization}
   Especially for the case of sparse graphs, it is useful to add a regularizer, which prevents the maximization routine to move towards high values of 
   $J_{ij}$ and $h_i$ without converging. We will adopt an $l_2$ regularization so that the Pseudolikelihood function (PLF) at site $i$  reads:
   \begin{equation}\label{eq:plf_i}
  {\cal L}_i = L_i 
  - \lambda \sum_{i \neq j} \left(J_{ij}^R\right)^2 - \lambda \sum_{i \neq j} \left(J_{ij}^I\right)^2  
   \end{equation}
   with $\lambda>0$.
   Note that the values of $\lambda$ have to be chosen arbitrarily, but not too large, in order not to overcome $L_i$.
   The standard implementation of the PLM consists in maximizing each ${\cal L}_i$, for $i=1\dots N$, separately. The expected values of the couplings are then:
   \begin{equation}
   \{ J_{i j}^*\}_{j\in \partial i} :=  \mbox{arg max}_{ \{ J_{ij} \}}
    \left[{\cal L}_i\right]
   \end{equation}
   In this way, we obtain two estimates for the coupling $J_{ij}$, one from maximization of ${\cal L}_i$, $J_{ij}^{(i)}$, and another one from ${\cal L}_j$, say $J_{ij}^{(j)}$.
    Since the original Hamiltonian of the $XY$ model is Hermitian, we know that the real part of the couplings is symmetric while the imaginary part is skew-symmetric. 
     The final estimate for $J_{ij}$ can then be obtained  averaging the two results:
   \begin{equation}\label{eq:symm}
   J_{ij}^{\rm inferred} = \frac{J_{ij}^{(i)} + \bar{J}_{ij}^{(j)}}{2} 
   \end{equation}
   where with $\bar{J}$ we indicate the complex conjugate.
   It is worth noting that the pseudolikelihood $L_i$, Eq. \eqref{eq:L_i}, is characterized by the
   following properties: (i) the normalization term of Eq.\eqref{eq:marginal_xy} can be
   computed analytically at odd with the {\em full} likelihood case that
   in general require a computational time which scales exponentially
   with the size of the systems; (ii) the $\ell_2$-regularized pseudolikelihood
   defined in Eq.\eqref{eq:plf_i} is strictly concave (i.e. it has a single
   maximizer)\cite{Ravikumar10}; (iii) it is consistent, i.e. if $M$ samples are
   generated by a model $P(\phi | J*)$ the maximizer tends to $J*$
   for $M\rightarrow\infty$\cite{besag1975}. Note also that (iii) guarantees that  
   $|J^{(i)}_{ij}-J^{(j)}_{ij}| \rightarrow 0$ for $M\rightarrow \infty$.
   In Secs. \ref{sec:res_reg}, \ref{sec:res_dec} 
   we report the results obtained and we analyze the performances of the PLM having taken the configurations from Monte-Carlo simulations of models whose details are known.

   \subsection{PLM with decimation}
\begin{figure}[t!]
   	\centering
   	\includegraphics[width=1\linewidth]{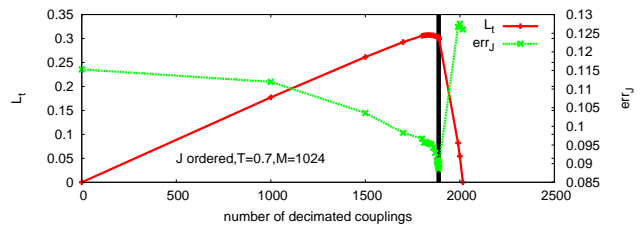}
   	\caption{The tilted likelyhood ${\cal L}_t$ curve and the reconstruction error vs the number of decimated couplings  for an ordered, real-valued J on 2D XY model with $N=64$ spins. The peak of ${\cal L}_t$ coincides with the dip of the error.}  
   	\label{Jor1-$t$PLF}
   \end{figure}

   Even though the PLM with $l_2$-regularization allows to dwell the inference towards the low temperature region and in the low sampling case with better performances that mean-field methods, in some situations some couplings are overestimated and not at all symmetric. Moreover, in the technique there is the bias of the $l_2$ regularizer.
   Trying to overcome these problems, Decelle and Ricci-Tersenghi introduced a new method \cite{Decelle14}, known as PLM + decimation: the algorithm maximizes the sum of the $L_i$,
   \begin{eqnarray}
    {\cal L}\equiv \frac{1}{N}\sum_{i=1}^N \mbox{L}_i
    \end{eqnarray}  
    and, then, it recursively set to zero couplings which are estimated very small. We expect that as long as we are setting to zero couplings that are unnecessary to fit the data, there should be not much changing on ${\cal L}$. Keeping on with decimation, a point is reached where ${\cal L}$ decreases abruptly indicating  that relevant couplings are being decimated and under-fitting is taking place.
   Let us define  by $x$  the fraction of non-decimated couplings. To have a quantitative measure for the halt criterion of the decimation process, a tilted ${\cal L}$ is defined as,
   \begin{eqnarray}
  \mathcal{L}_t &\equiv& \mathcal{L}  - x \mathcal{L}_{\textup{max}} - (1-x) \mathcal{L}_{\textup{min}} \label{$t$PLF} 
   \end{eqnarray}
   where 
   \begin{itemize}
   \item $\mathcal{L}_{\textup{min}}$ is the pseudolikelyhood of a model with independent variables. In the XY case: $\mathcal{L}_{\textup{min}}=-\ln{2 \pi}$.
   \item
   $\mathcal{L}_{\textup{max}}$ is the pseudolikelyhood in the fully-connected model and it is maximized over all the $N(N-1)/2$ possible couplings. 
   \end{itemize}
   At the first step, when $x=1$, $\mathcal{L}$ takes value $\mathcal{L}_{\rm max}$ and  $\mathcal{L}_t=0$. On the last step, for an empty graph, i.e., $x=0$, $\mathcal{L}$ takes the value $\mathcal{L}_{\rm min}$ and, hence, again $\mathcal{L}_t =0$. 
   In the intermediate steps, during the decimation procedure, as $x$ is decreasing from $1$ to $0$, one observes firstly that $\mathcal{L}_t$ increases linearly and, then, it displays an abrupt decrease indicating that from this point on relevant couplings are being decimated\cite{Decelle14}. In Fig. \ref{Jor1-$t$PLF} we give an instance of this behavior for the 2D short-range XY model with ordered couplings. We notice that the maximum point of $\mathcal{L}_t$ coincides with the minimum point of the reconstruction error, the latter defined as 
   \begin{eqnarray}\label{eq:errj}
   \mbox{err}_J \equiv \sqrt{\frac{\sum_{i<j} (J^{\rm inferred}_{ij} -J^{\rm true}_{ij})^2}{N(N-1)/2}} \label{err}
   \end{eqnarray}
  We stress that the ${\cal L}_t$ maximum is obtained ignoring the underlying graph, while the err$_J$ minimum can be evaluated once the true graph has been reconstructed. 
   
   In the next sections we will show the results obtained on the  $XY$ model analyzing the performances of the two methods and comparing them also with a mean-field method \cite{Tyagi15}.
   
    
   \section{Inferred couplings with PLM-$l_2$}
    \label{sec:res_reg}
    \subsection{$XY$ model with real-valued couplings}
    
    In order to obtain the vector of couplings, $J_{ij}^{\rm inferred}$ the function $-\mathcal{L}_i$ is minimized through the vector of derivatives ${\partial \mathcal{L}_i}/\partial J_{ij}$. The process is repeated for all the couplings obtaining then a fully connected adjacency matrix. The results here presented are obtained with $\lambda = 0.01$.
   For the minimization we have used the MATLAB routine \emph{minFunc\_2012}\cite{min_func}. 
      
      \begin{figure}[t!]
    	\centering
    	\includegraphics[width=1\linewidth]{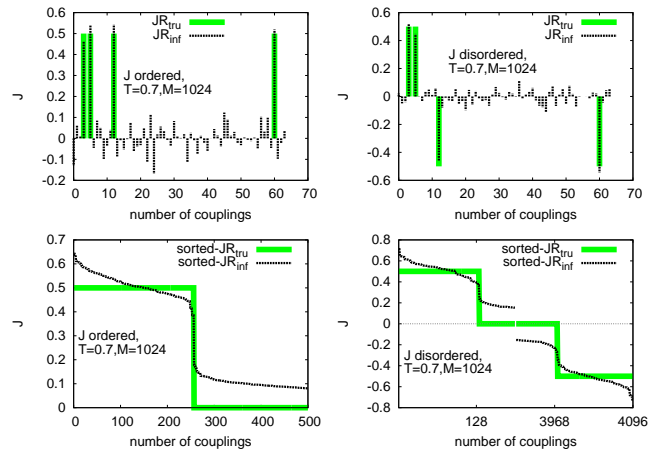}
    	\caption{Top panels: instances of single site coupling reconstruction for the case of $N=64$ XY spins on a 2D lattice with ordered $J$ (left column) and bimodal distributed $J$ (right column). 
    	Bottom panels: sorted couplings.}
    	\label{PL-Jor1}
    \end{figure}

To produce the data by means of numerical Monte Carlo simulations a system with $N=64$  spin variables  is considered on a deterministic 2D lattice with periodic boundary conditions. 
Each spin has then connectivity $4$, i.e., we expect to infer an adjacency matrix with $N c = 256$ couplings different from zero. 
The dynamics of the simulated model is based on the Metropolis algorithm and parallel tempering\cite{earl05} is used to speed up the thermalization of the system.
The thermalization is tested looking at the average energy over logarithmic time windows and
the acquisition of independent configurations 
starts only after the system is well thermalized.

   For the values of the couplings we considered two cases: an ordered case, indicated in the figure as  $J$ ordered (e.g., left column of Fig. \ref{PL-Jor1}) where the couplings can take values $J_{ij}=0,J$, with $J=1$, 
   and a quenched disordered case, indicated in the figures as  $J$ disordered (e.g., right column of Fig. \ref{PL-Jor1})
   where the couplings can take also  negative values, i.e., 
    $J_{ij}=0,J,-J$, with a certain probability.  The results here presented were obtained with bimodal distributed $J$s: 
    $P(J_{ij}=J)=P(J_{ij}=-J)=1/2$.  The performances of the PLM have shown not to depend on $P(J)$. 
    We recall that in Sec. \ref{sec:plm} we used the temperature-rescaled notation, i.e., $J_{ij}$ stands for $J_{ij}/T$. 
   
    To analyze the performances of the PLM, in Fig. \ref{PL-Jor1} the inferred couplings, $\mathbb{J}^R_{\rm inf}$, are shown on top of the original couplings,  $\mathbb{J}^R_{\rm true}$.
     The first figure (from top) in the left column shows  the $\mathbb{J}^R_{\rm inf}$ (black) and the $\mathbb{J}^R_{\rm tru}$ (green) for a given spin 
     at temperature $T/J=0.7$ and number of samples  $M=1024$. PLM appears to reconstruct the correct couplings, though zero couplings are always given a small inferred non-zero value. 
     In the left column of Fig.  \ref{PL-Jor1},  both the $\mathbb{J}^R_{\rm{inf}}$ and the $\mathbb{J}^R_{\rm{tru}}$ are sorted in decreasing order and plotted on top of each other. 
     We can clearly see that $\mathbb{J}^R_{\rm inf}$ reproduces the expected step function. Even though the jump is smeared, the difference between inferred couplings corresponding to the set of non-zero couplings 
     and to the set of zero couplings can be clearly appreciated.
     Similarly, the plots in the right column of Fig. \ref{PL-Jor1} show the results obtained for the case with  bimodal disordered couplings, for the same working temperature and number of samples. 
     In particular, note that the algorithm infers half positive and half negative couplings, as expected.

\begin{figure}
\centering
\includegraphics[width=1\linewidth]{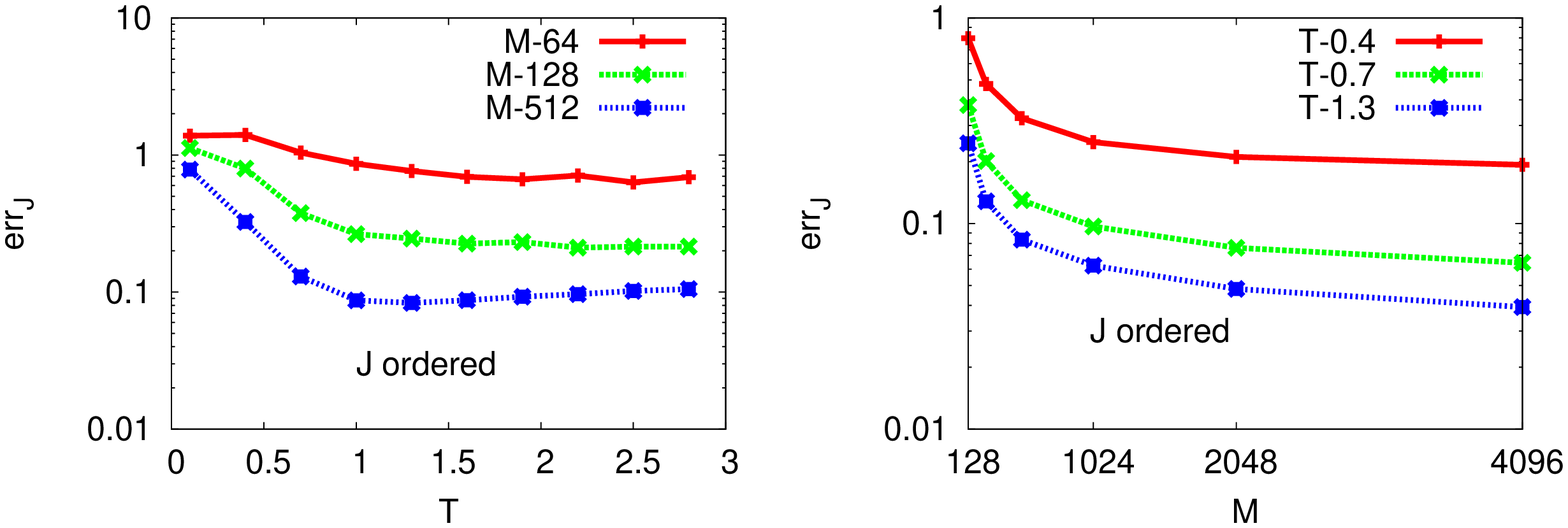}
\caption{Reconstruction error $\mbox{err}_J$, cf. Eq. (\ref{eq:errj}), plotted as a function of temperature (left) for three values of the number of samples $M$ and  as a function $M$ (right) for three values of temperature in the ordered system, i.e., $J_{ij}=0,1$. 
The system size is $N=64$.}
\label{PL-err-Jor1}
\end{figure}

In order to analyze the effects of the number of samples and of the temperature regimes, we plot in Fig. \ref{PL-err-Jor1} the reconstruction error, Eq. (\ref{err}), as a function of temperature for three different sample sizes $M=64,128$ and $512$. 
The error is seen to sharply rise al low temperature, incidentally, in the ordered case, for  $T<T_c \sim 0.893$, which is the Kosterlitz-Thouless transition temperature of the 2XY model\cite{Olsson92}. 
 However, we can see that if only $M=64$ samples are considered, $\mbox{err}_J$ remains high independently on the working temperature. 
 In the right plot of Fig. \ref{PL-err-Jor1},  $\mbox{err}_J$ is plotted as a function of $M$ for three different working temperatures $T/J=0.4,0.7$ and $1.3$. As we expect, 
 $\mbox{err}_J$  decreases as $M$ increases. This effect was observed also with mean-field inference techniques on the same model\cite{Tyagi15}.

\begin{figure}[t!]
	\centering
	\includegraphics[width=1\linewidth]{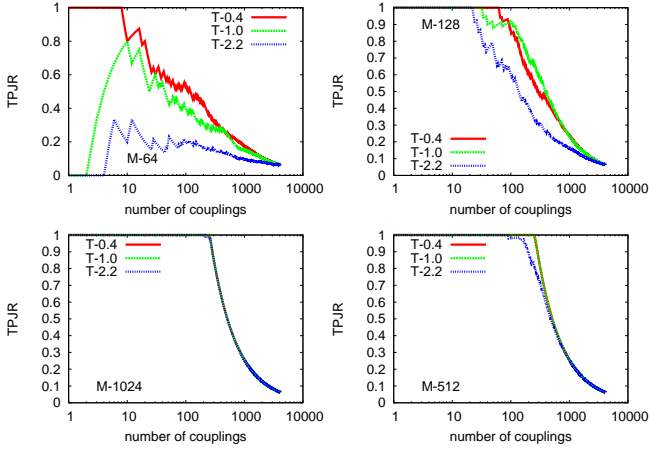}
	\caption{TP  curves  for 2D short-range ordered $XY$ model with $N=64$ spins at three different values of $T/J$ with increasing - clockwise from top - $M$.}
	\label{PL-varTP-Jor1} 
\end{figure}

To better understand the performances of the algorithms, in Fig. \ref{PL-varTP-Jor1} we show several True Positive (TP) curves obtained for various values of $M$ at three different temperatures $T$. As $M$ is large and/or temperature is not too small,  we are able to reconstruct correctly all the couplings present in the system (see bottom plots).
The True Positive curve displays how many times the inference method finds a true link of the original network  as a function of the index of the vector of sorted absolute value of reconstructed couplings $J_{ij}^{\rm inf}$. 
The index $n_{(ij)}$ represents the related spin couples $(ij)$. The TP curve is obtained as follows: 
first  the values $|J^{\rm inf}_{ij}|$ are sorted in descending order and  the spin pairs $(ij)$ are ordered according to the sorting position of $|J^{\rm inf}_{ij}|$. Then,
     	a cycle over the ordered set of pairs $(ij)$, indexed by $n_{(ij)}$, is performed, comparing with the original network coupling $J^{\rm true}_{ij}$ and verifying whether it is zero or not. The true positive curve is computed as
\begin{equation}
\mbox{TP}[n_{(ij)}]= \frac{\mbox{TP}\left[n_{(ij)}-1\right] (n_{ij}-1)+ 1 -\delta_{J^{\rm true}_{ij},0}}{n_{(ij)}}
\end{equation}
As far as $J^{\rm true}_{ij} \neq 0$, TP$=1$. As soon as the true coupling of a given $(ij)$ couple in the sorted list is zero, the TP curve departs from one. 
In our case, where the connectivity per spin of the original system is  $c=4$ and there are $N=64$ spins, we know that we will have  $256$ non-zero couplings.  
     	If the inverse problem is successful, hence, we expect a steep decrease of the TP curve when  $n_{ij}=256$ is overcome.

In Fig. \ref{PL-varTP-Jor1}
it is  shown that,  almost independently of $T/J$, the TP score improves as $M$ increases. Results are plotted for three different temperatures, $T=0.4,1$ and $2.2$, with increasing number of samples $M = 64, 128,512$ and $1024$ (clockwise). 
We can clearly appreciate the improvement in temperature if the size of the data-set is not very large: for small $M$, $T=0.4$ performs better. 
When $M$ is high enough (e.g., $M=1024$), instead, the TP curves do not appear to be strongly influenced by the temperature.

\subsection{$XY$ model with complex-valued couplings}
\begin{figure}[t!]
	\centering
\includegraphics[width=1\linewidth]{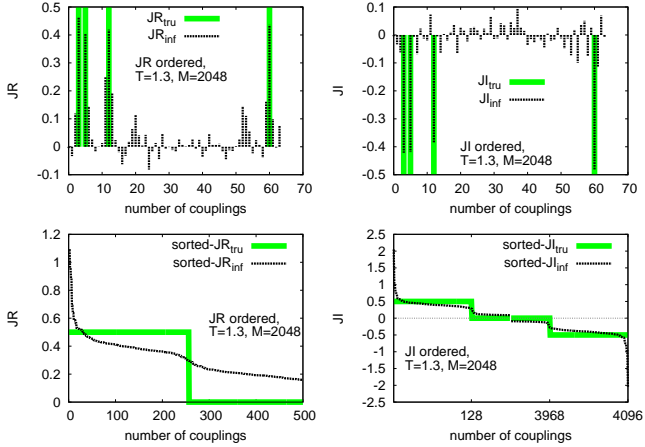}
	\caption{Results related to the ordered complex XY model with $N=64$ spins on a  2D lattice. Top: instances of  single site reconstruction for the real, JR (left column), and
		the imaginary, JI (right column), part of $J_{ij}$. Bottom: sorted values of JR (left) and JI (right).}
	\label{PL-Jor3}
\end{figure}

For the complex $XY$ we have to contemporary  infer $2$ apart coupling matrices,  $J^R_{i j}$ and $J^I_{i j}$.  As before, a system of $N=64$ spins is considered on a 2D lattice.
For the couplings we have considered both ordered and bimodal disordered cases.
In Fig. \ref{PL-Jor3}, a single row of the matrix $J$ (top) and the whole sorted couplings (bottom) are displayed for the ordered model (same legend as in Fig. \ref{PL-Jor1}) for the real, $J^R$ (left column), and the imaginary part, $J^I$.  

  \section{PLM with Decimation}
 \label{sec:res_dec}

\begin{figure}[t!]
   	\centering
   	\includegraphics[width=1\linewidth]{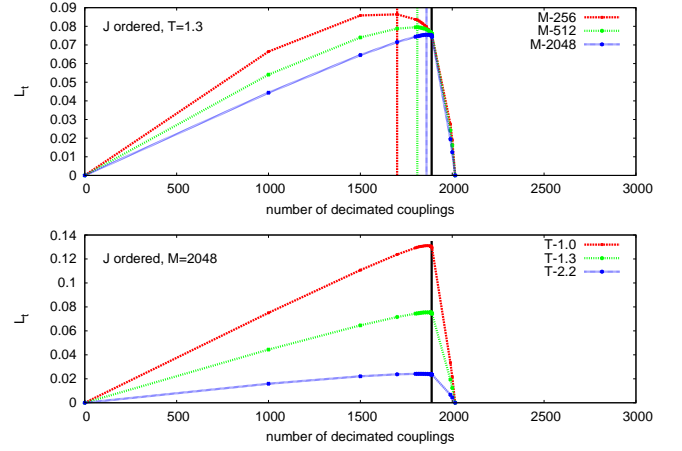}
    	\caption{Tilted Pseudolikelyhood, ${\cal L}_t$, plotted as a function of decimated couplings. Top: Different ${\cal L}_t$ curves obtained for different values of $M$ plotted on top of each other. Here $T=1.3$. The black line indicates the expected number of decimated couplings, $x^*=(N (N-1) - N c)/2=1888$. As we can see, as $M$ increases, the maximum point of ${\cal L}_t$ approaches $x^*$. Bottom: Different ${\cal L}_t$ curves obtained for different values of T with $M=2048$. We can see that, with this value of $M$, no differences can be appreciated on the maximum points of the different  ${\cal L}_t$ curves.}
   	\label{var-$t$PLF}
   \end{figure}

\begin{figure}[t!]
    \centering
    \includegraphics[width=1\linewidth]{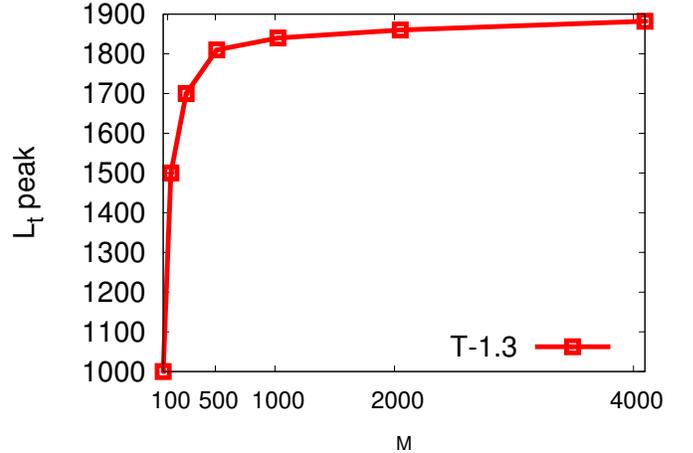}
    	\caption{Number of most likely decimated couplings, estimated by the maximum point of $\mathcal{L}_t$, as a function of the number of samples $M$. We can clearly see that the maximum point of $\mathcal{L}_t$ tends toward $x^*$, which is the right expected number of zero couplings in the system.}  
    	\label{PLF_peak_statistics}
    \end{figure}
      
\begin{figure}[t!]
    	\centering
      	\includegraphics[width=1\linewidth]{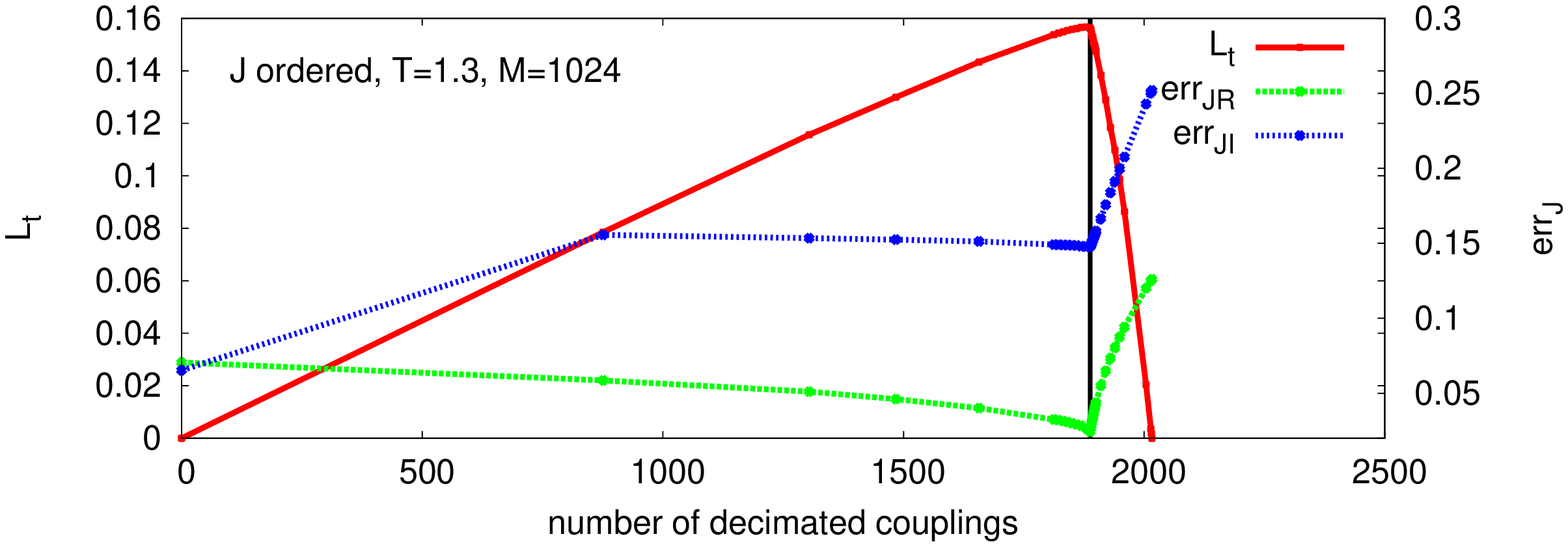}
      	\caption{Tilted Pseudolikelyhood, ${\cal L}_t$, plotted with the reconstruction errors for the XY model with $N=64$ spins on a 2D lattice. These results refer to the case of  ordered and complex valued couplings. The full (red) line indicates ${\cal L}_t$. The dashed (green) 
      		and the dotted (blue) lines show the reconstruction errors (Eq. \eqref{eq:errj}) obtained for the real and the imaginary couplings respectively. We can see that both ${\rm err_{JR}}$ and ${\rm err_{JI}}$ have a minimum at $x^*$.}
          	\label{fig:Lt_complex}
    \end{figure}

      For the ordered real-valued XY model we show in Fig. \ref{var-$t$PLF}, top panel, the outcome on the tilted pseudolikelyhood, $\mathcal{L}_t$ Eq. \eqref{$t$PLF}, of the progressive decimation: from a fully connected lattice  down to an empty lattice. The figure shows the behaviour of $\mathcal{L}_t$ for three different data sizes $M$. A clear data size dependence of the maximum point of  $\mathcal{L}_t$, signalling the most likely value for decimation, is shown. For small $M$ the most likely number of  couplings is overestimated and for increasing $M$ it tends to the true value, as displayed in Fig. \ref{PLF_peak_statistics}.  In the bottom panel of Fig. \ref{var-$t$PLF} we display instead different 
 $\mathcal{L}_t$ curves obtained for three different values of $T$.
  Even though the values of $\mathcal{L}_t$ decrease with increasing temperature, the value of the most likely number of decimated couplings appears to be quite independent on $T$ with $M=2048$ number of samples.
In Fig. \ref{fig:Lt_complex} we eventually display the tilted pseudolikelyhood for a 2D network with complex valued ordered couplings, where the decimation of the real and imaginary coupling matrices proceeds in parallel, that is, 
when a real coupling is small enough to be decimated its imaginary part is also decimated, and vice versa.
One can see that though the apart errors for the real and imaginary parts are different in absolute values, they display the same dip, to be compared with the maximum point of $\mathcal{L}_t$.
     
\begin{figure}[t!]
   	\centering
   	\includegraphics[width=1\linewidth]{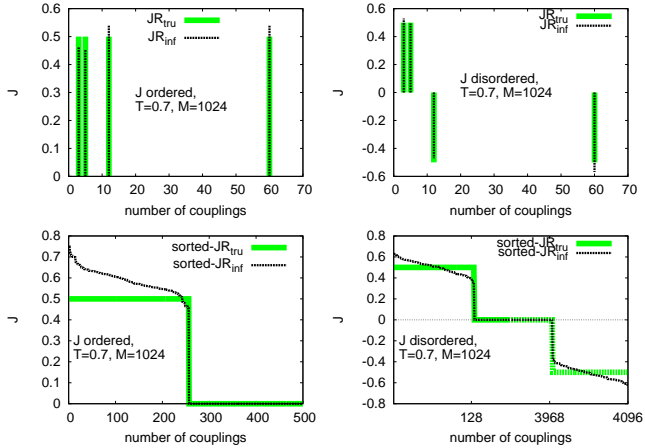}
   	\caption{XY model on a 2D lattice with $N=64$ sites and real valued couplings. The graphs show the inferred (dashed black lines) and true couplings (full green lines) plotted on top of each other. The left and right columns refer to the
   		 cases of ordered and bimodal disordered couplings, respectively. Top figures: single site reconstruction, i.e., one row of the matrix $J$. Bottom figures: couplings are plotted sorted in descending order.}  
   	\label{Jor1_dec}
   \end{figure}
   
\begin{figure}[t!]
    	\centering
    	\includegraphics[width=1\linewidth]{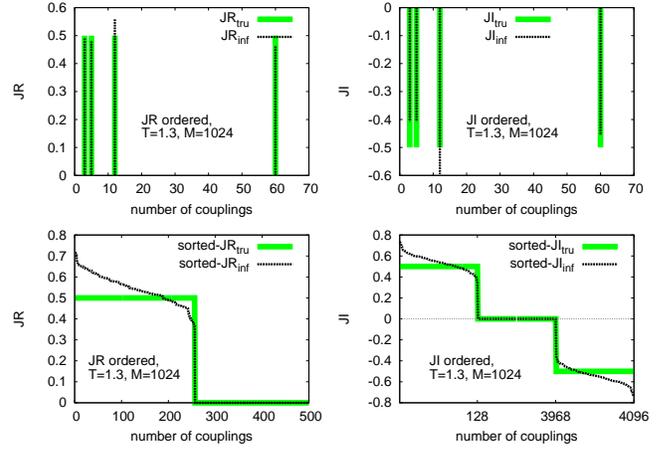}
    	\caption{XY model on a 2D lattice with $N=64$ sites and ordered complex-valued couplings.
    		The inferred and true couplings are plotted on top of each other. The left and right columns show the real and imaginary parts, respectively, of the couplings. Top figures refer to a single site reconstruction, i.e., one row of the matrix $J$. Bottom figures report the couplings sorted in descending order.}
    	\label{Jor3_dec}
    \end{figure}


       \begin{figure}[t!]
     	\centering
     	\includegraphics[width=1\linewidth]{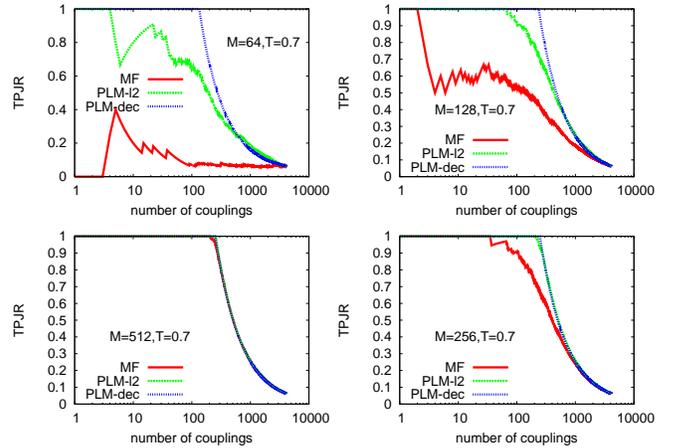}
     	\caption{True Positive curves obtained with the three techniques: PLM with decimation, (blue) dotted line,  PLM with $l_2$ regularization, (greed) dashed line, and mean-field, (red) full line.  These results refer to real valued ordered couplings with $N=64$ spins on a 2D lattice. The temperature is here $T=0.7$ while the four graphs refer to different sample sizes: $M$ increases clockwise.}
     	\label{MF_PL_TP}
     \end{figure}
    
    \begin{figure}[t!]
    	\centering
    	\includegraphics[width=1\linewidth]{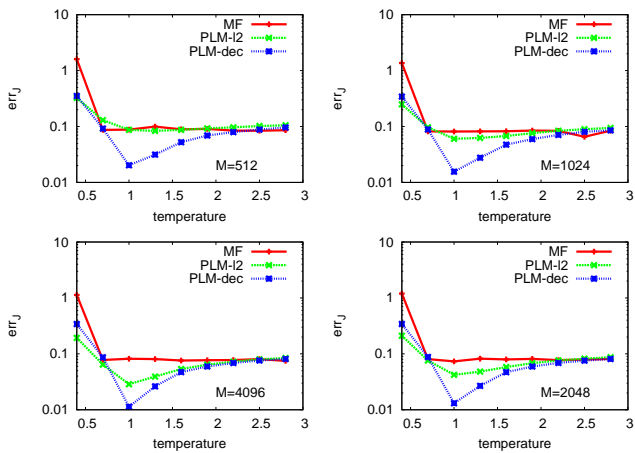}
    	\caption{Variation of reconstruction error, ${\rm err_J}$, with respect to temperature as obtained with the three different techniques, see Fig. \ref{MF_PL_TP}, for four different sample size:  clockwise from top   $M=512,1024, 2048$ and $4096$.} 
    	\label{MF_PL_err}
    \end{figure}
    
     Once the most likely network has been identified through the decimation procedure, we perform the same analysis displayed in Fig. \ref{Jor1_dec}  for ordered and then quenched disordered real-valued couplings
and in Fig. \ref{Jor3_dec} for  complex-valued ordered couplings.  In comparison to the results shown in Sec. \ref{sec:res_reg},
  the PLM with decimation leads to rather cleaner results. In Figs. \ref{MF_PL_err} and \ref{MF_PL_TP} we compare the performances of the PLM with decimation in respect to ones of the PLM with $l_2$-regularization. These two techniques are also analysed in respect to a mean-field technique previously implemented on the same XY systems\cite{Tyagi15}.
  
    For what concerns the network of connecting links, in Fig. \ref{MF_PL_TP} we compare the TP curves obtained with the three techniques. The results refer to the case of ordered and real valued couplings, but similar behaviours were obtained for the other cases analysed. 
  The four graphs are related to different sample sizes, with $M$ increasing clockwise. When $M$ is high enough, all techniques reproduce the true network. 
  However, for lower values of $M$ the performances of the PLM with $l_2$ regularization and with decimation drastically overcome those ones of the previous mean field technique. 
  In particular, for $M=256$ the PLM techniques still reproduce the original network while the mean-field method fails to find more than half of the couplings. 
  When $M=128$, the network is clearly reconstructed only through the PLM with decimation while the PLM with $l_2$ regularization underestimates the couplings. 
  Furthermore, we notice that the PLM method with decimation is able to clearly infer the network of interaction even when $M=N$ signalling that it could be considered also in the under-sampling regime $M<N$.  
In Fig. \ref{MF_PL_err} we compare the temperature behaviour of the reconstruction error.
In can be observed that for all temperatures and for all sample sizes  the reconstruction error, ${\rm err_J}$, (plotted here in log-scale) obtained with the PLM+decimation is always smaller than 
that one obtained with the other techniques. The temperature behaviour of ${\rm err_J}$ agrees with the one already observed for Ising spins in \cite{Nguyen12b} and for XY spins  in \cite{Tyagi15} with a mean-field approach:  ${\rm err_J}$ displays a minimum around $T\simeq 1$ and then it increases for very lower $T$; however,
 the error obtained with the PLM with decimation is several times smaller  than the error estimated by the other methods.

     \section{Conclusions}
     \label{sec:conc}

Different statistical inference methods have been applied to the inverse problem of the XY model.
After a short review of techniques based on pseudo-likelihood and their formal generalization to the model we have tested their performances against data generated by means of Monte Carlo numerical simulations of known instances
with diluted, sparse, interactions.

The main outcome is that the best performances are obtained by means of the  pseudo-likelihood method combined with decimation. Putting to zero (i.e., decimating) very weak bonds, this technique turns out to be very precise for  problems whose real underlying interaction network is sparse, i.e., the number of couplings per variable does not scale with number of variables.
The PLM + decimation method is compared to the PLM + regularization method, with $\ell_2$ regularization and to a mean-field-based method. The behavior of the quality of the network reconstruction is analyzed by looking at the overall sorted couplings and at the single site couplings, comparing them with the real network, and at the true positive curves in all three approaches. In the PLM +decimation method, moreover, the identification of the number of decimated bonds at which the tilted pseudo-likelihood is maximum allows for a precise estimate of the total number of bonds. Concerning this technique, it is also shown that the network with the most likely number of bonds is also the one of least reconstruction error, where not only the prediction of the presence of a bond is estimated but also its value.

The behavior of the inference quality in temperature and in the size of data samples is also investigated, basically confirming the low $T$ behavior hinted by Nguyen and Berg \cite{Nguyen12b} for the Ising model. In temperature, in particular, the reconstruction error curve displays a minimum at a low temperature, close to the critical point in those cases in which a critical behavior occurs, and a sharp increase as temperature goes to zero. The decimation method, once again, appears to enhance this minimum of the reconstruction error of almost an order of magnitude with respect to other methods.
 
The techniques displayed and the results obtained in this work can be of use in any of the many systems whose theoretical representation is given by Eq. \eqref{eq:HXY} or Eq. \eqref{eq:h_im}, some of which are recalled in Sec. \ref{sec:model}. In particular, a possible application can be the field of light waves propagation through random media and the corresponding problem of the  reconstruction of an object seen through an opaque medium or a disordered optical fiber \cite{Vellekoop07,Vellekoop08a,Vellekoop08b, Popoff10a,Akbulut11,Popoff11,Yilmaz13,Riboli14}.

 	\bibliography{Lucabib}

 \end{document}